\input harvmac
\let\includefigures=\iftrue
\let\useblackboard=\iftrue
\newfam\black

\includefigures
\message{If you do not have epsf.tex (to include figures),}
\message{change the option at the top of the tex file.}
\input epsf
\def\figin{\epsfcheck\figin}\def\figins{\epsfcheck\figins}
\def\epsfcheck{\ifx\epsfbox\UnDeFiNeD
\message{(NO epsf.tex, FIGURES WILL BE IGNORED)}
\gdef\figin##1{\vskip2in}\gdef\figins##1{\hskip.5in}
\else\message{(FIGURES WILL BE INCLUDED)}%
\gdef\figin##1{##1}\gdef\figins##1{##1}\fi}
\def\DefWarn#1{}
\def\figinsert{\goodbreak\midinsert}
\def\ifig#1#2#3{\DefWarn#1\xdef#1{fig.~\the\figno}
\writedef{#1\leftbracket fig.\noexpand~\the\figno}%
\figinsert\figin{\centerline{#3}}\medskip\centerline{\vbox{
\baselineskip12pt\advance\hsize by -1truein
\noindent\footnotefont{\bf Fig.~\the\figno:} #2}}
\endinsert\global\advance\figno by1}
\else
\def\ifig#1#2#3{\xdef#1{fig.~\the\figno}
\writedef{#1\leftbracket fig.\noexpand~\the\figno}%
\global\advance\figno by1} \fi
%
\def\wt{\widetilde}

\def\id{{1 \kern-.28em {\rm l}}}

\def\K3{{\bf K3}}
\def\journal#1&#2(#3){\unskip, \sl #1\ \bf #2 \rm(19#3) }
\def\andjournal#1&#2(#3){\sl #1~\bf #2 \rm (19#3) }

\def\bar{\overline}

\def\ie{{\it i.e.}}
\def\eg{{\it e.g.}}

\def\tilde{\widetilde}

\def\frac#1#2{{#1\over#2}}

\def\half{\frac12}

\def\inbar{\,\vrule height1.5ex width.4pt depth0pt}
\def\IC{\relax\hbox{$\inbar\kern-.3em{\rm C}$}}
\def\IR{\relax{\rm I\kern-.18em R}}
\def\IP{\relax{\rm I\kern-.18em P}}

%
%

%
\catcode`\@=11
\def\slash#1{\mathord{\mathpalette\c@ncel{#1}}}
\overfullrule=0pt

\def\DD{{\cal D}}

\def\KK{{\cal K}}
\def\LL{{\cal L}}

\def\OO{{\cal O}}

\def\SS{{\cal S}}

\def\VV{{\cal V}}
\def\WW{{\cal W}}

\def\e{\epsilon}

\def\underrel#1\over#2{\mathrel{\mathop{\kern\z@#1}\limits_{#2}}}

\catcode`\@=12


%

\def\det{{\rm det}}

\def\det{{\rm det}}


\def\Phibar{{\bar\Phi}}

\def\vbar{{\bar v}}
\def\wbar{{\overline w}}

\lref\ElitzurFH{
  S.~Elitzur, A.~Giveon and D.~Kutasov,
  ``Branes and N = 1 duality in string theory,''
  Phys.\ Lett.\  B {\bf 400}, 269 (1997)
  [arXiv:hep-th/9702014].
}

\lref\ElitzurHC{
  S.~Elitzur, A.~Giveon, D.~Kutasov, E.~Rabinovici and A.~Schwimmer,
  ``Brane dynamics and N = 1 supersymmetric gauge theory,''
  Nucl.\ Phys.\  B {\bf 505}, 202 (1997)
  [arXiv:hep-th/9704104].
}

\lref\GiveonSR{
  A.~Giveon and D.~Kutasov,
  ``Brane dynamics and gauge theory,''
  Rev.\ Mod.\ Phys.\  {\bf 71}, 983 (1999)
  [arXiv:hep-th/9802067].
}

\lref\EssigKZ{
  R.~Essig, J.~F.~Fortin, K.~Sinha, G.~Torroba and M.~J.~Strassler,
  ``Metastable supersymmetry breaking and multitrace deformations of SQCD,''
  arXiv:0812.3213 [hep-th].
}

\lref\IntriligatorDD{
  K.~A.~Intriligator, N.~Seiberg and D.~Shih,
  ``Dynamical SUSY breaking in meta-stable vacua,''
  JHEP {\bf 0604}, 021 (2006)
  [arXiv:hep-th/0602239].
}

\lref\GiveonEW{
  A.~Giveon and D.~Kutasov,
  ``Stable and Metastable Vacua in Brane Constructions of SQCD,''
  JHEP {\bf 0802}, 038 (2008)
  [arXiv:0710.1833 [hep-th]].
}

\lref\GiveonEF{
  A.~Giveon and D.~Kutasov,
  ``Stable and Metastable Vacua in SQCD,''
  Nucl.\ Phys.\  B {\bf 796}, 25 (2008)
  [arXiv:0710.0894 [hep-th]].
}

\lref\GiveonFK{
  A.~Giveon and D.~Kutasov,
  ``Gauge symmetry and supersymmetry breaking from intersecting branes,''
  Nucl.\ Phys.\  B {\bf 778}, 129 (2007)
  [arXiv:hep-th/0703135].
}

\lref\NambuTP{Y.~Nambu and G.~Jona-Lasinio,
``Dynamical Model Of Elementary Particles Based On An Analogy With
Superconductivity. I,''Phys.\ Rev.\  {\bf 122}, 345 (1961).
}

\lref\AntonyanVW{
  E.~Antonyan, J.~A.~Harvey, S.~Jensen and D.~Kutasov,
  ``NJL and QCD from string theory,''
  arXiv:hep-th/0604017.
}

\lref\CallanAT{
  C.~G.~Callan, J.~A.~Harvey and A.~Strominger,
  ``Supersymmetric string solitons,''
  arXiv:hep-th/9112030.
}

\lref\MarolfVF{
  D.~Marolf, L.~Martucci and P.~J.~Silva,
  ``Actions and fermionic symmetries for D-branes in bosonic backgrounds,''
  JHEP {\bf 0307}, 019 (2003)
  [arXiv:hep-th/0306066].
}

\lref\WessCP{
  J.~Wess and J.~Bagger,
  ``Supersymmetry and supergravity,''
{\it  Princeton, USA: Univ. Pr. (1992) 259 p.}
}

\lref\BeasleyDC{
  C.~Beasley, J.~J.~Heckman and C.~Vafa,
  JHEP {\bf 0901}, 058 (2009)
  [arXiv:0802.3391 [hep-th]].
}

\lref\BeasleyKW{
  C.~Beasley, J.~J.~Heckman and C.~Vafa,
  JHEP {\bf 0901}, 059 (2009)
  [arXiv:0806.0102 [hep-th]].
}

\lref\OoguriBG{
  H.~Ooguri and Y.~Ookouchi,
  ``Meta-stable supersymmetry breaking vacua on intersecting branes,''
  Phys.\ Lett.\  B {\bf 641}, 323 (2006)
  [arXiv:hep-th/0607183].
}

\lref\FrancoHT{
  S.~Franco, I.~Garcia-Etxebarria and A.~M.~Uranga,
  ``Non-supersymmetric meta-stable vacua from brane configurations,''
  JHEP {\bf 0701}, 085 (2007)
  [arXiv:hep-th/0607218].
}

\lref\BenaRG{
  I.~Bena, E.~Gorbatov, S.~Hellerman, N.~Seiberg and D.~Shih,
  ``A note on (meta)stable brane configurations in MQCD,''
  JHEP {\bf 0611}, 088 (2006)
  [arXiv:hep-th/0608157].
}

\lref\PolchinskiRR{
  J.~Polchinski,
  ``String theory. Vol. 2: Superstring theory and beyond,''
{\it  Cambridge, UK: Univ. Pr. (1998) 531 p.}
}

\lref\GiudiceBP{
  G.~F.~Giudice and R.~Rattazzi,
  ``Theories with gauge-mediated supersymmetry breaking,''
  Phys.\ Rept.\  {\bf 322}, 419 (1999)
  [arXiv:hep-ph/9801271].
}

\lref\AganagicPE{
  M.~Aganagic, C.~Popescu and J.~H.~Schwarz,
  ``D-brane actions with local kappa symmetry,''
  Phys.\ Lett.\  B {\bf 393}, 311 (1997)
  [arXiv:hep-th/9610249].
}

\lref\BergshoeffTU{
  E.~Bergshoeff and P.~K.~Townsend,
  ``Super D-branes,''
  Nucl.\ Phys.\  B {\bf 490}, 145 (1997)
  [arXiv:hep-th/9611173].
}

\Title{}
{\vbox{\centerline{D-Terms and Supersymmetry Breaking from Branes}
\bigskip
\centerline{}
}}
\bigskip

\centerline{\it Amit Giveon\foot{Permanent address: Racah
Institute of Physics, The Hebrew University, Jerusalem 91904, Israel.},
David Kutasov, Jock McOrist and Andrew B. Royston}
\bigskip
\smallskip
\centerline{EFI and Department of Physics, University of
Chicago} \centerline{5640 S. Ellis Av., Chicago, IL 60637, USA }

\smallskip

\vglue .3cm

\bigskip

\let\includefigures=\iftrue
\bigskip
\noindent
Metastable supersymmetry breaking configurations of $D$-branes and $NS5$-branes in string theory often
owe their existence to classical gravitational interactions between the branes. We show that in the effective
theory of the light fields, these interactions give rise to a non-canonical K\"ahler potential and other
D-terms. String theory provides a UV completion in which these non-renormalizable terms can be computed.
We use these observations to clarify the relation between the phase structure of ISS-type models and their
brane realizations.

\bigskip

\Date{}

\newsec{Introduction}

In the last few years there has been some work on metastable supersymmetry
breaking vacua in supersymmetric field theories, following the observation
of ISS \IntriligatorDD\ that such vacua may be rather generic. The particular
example studied in \IntriligatorDD\ -- massive supersymmetric QCD (SQCD) in the
free magnetic phase -- reduces in the infrared to an effective Wess-Zumino (WZ)
model for the light fields, which captures the phase structure of the theory.
The parameters of this ``macroscopic'' WZ model are determined in terms of
those of the  underlying ``microscopic'' SQCD, which provides an ultraviolet (UV)
completion.

Extending the discussion of \IntriligatorDD\ to string theory is interesting since
it gives rise to a more general class of UV completions, which may exhibit new phenomena.
This can be done \refs{\OoguriBG\FrancoHT\BenaRG-\GiveonFK} by using the well known
realization of SQCD  as the low energy theory on  intersecting $D$-branes and
$NS5$-branes (see \GiveonSR\ for a review).  In the brane description, the light fields
of \IntriligatorDD\ correspond to open strings stretched between various $D$-branes.
Their low energy dynamics is described by the field theory studied in \IntriligatorDD,
with corrections that depend on the parameters of the brane configuration.

The WZ model of  \IntriligatorDD\ contains pseudo-moduli (\ie\ complex massless scalar
fields $\Phi$ with a classically flat potential), whose expectation values parameterize
a moduli space of SUSY breaking vacua.  This moduli space is lifted by one loop
effects, which give a mass to $\Phi$  and stabilize it at the origin of pseudo-moduli space.
In the brane description, when all the distances between the branes are large, the one loop
effects of \IntriligatorDD\ are in general small and the dominant contribution to the  potential
on pseudo-moduli space is due to the classical gravitational attraction between $D$-branes
and $NS5$-branes \GiveonFK. It leads to the same qualitative outcome, stabilizing $\Phi$ at
the origin of pseudo-moduli space, but the origin of this stabilization is different from
that of \IntriligatorDD.

A natural question, which was not addressed in \GiveonFK, concerns the interpretation of
the above gravitational attraction in the low energy WZ model. The
purpose of this paper is to fill this gap. We will see that it gives rise to a non-trivial
K\"ahler potential and other, higher order, D-terms. From the low energy point of view,
these terms correspond to non-renormalizable interactions, and thus depend on the UV
completion of the WZ model. The embedding in string theory provides
such a completion, and determines all these terms.

One of the main motivations for this work is to better understand the generalized
ISS model studied in \refs{\GiveonEF\GiveonEW-\EssigKZ}. In field theory this model
is obtained by deforming the superpotential of $\Phi$. The metastable vacuum structure
of the resulting WZ model was analyzed in \GiveonEF\ and \EssigKZ\ (which
corrected a mistake in \GiveonEF). The corresponding brane system, which is obtained
from that of \refs{\OoguriBG\FrancoHT\BenaRG-\GiveonFK} by rotating some of the branes,
was analyzed in \GiveonEW. While the qualitative phase structures in field and string
theory agree, some important aspects are different. In particular, the
brane construction gives rise to metastable vacua not seen in the field theory.
Interestingly, these vacua are the more phenomenologically promising ones.

The interpretation of the vacuum structure of the brane system in terms of the low energy
effective field theory, that we will describe below, helps clarify the situation. The
additional vacua that are found in string theory, but not in the WZ model, occur when
the effective coupling at the UV cutoff scale exceeds a certain critical value. The
resulting non-renormalizable field theory is strongly coupled at high energies and
requires a UV completion. Such a completion is provided by string theory.

This situation is reminiscent of the Nambu-Jona-Lasinio (NJL) model \NambuTP, where the
vacuum spontaneously breaks chiral symmetry due to an attractive non-renormalizable
four-Fermi interaction. Symmetry breaking occurs when the four Fermi coupling at the UV
cutoff scale exceeds a certain critical value. A brane system in string theory provides a UV
complete theory with closely related dynamics \AntonyanVW.

The plan of this paper is the following. In section 2 we discuss the ISS model and its brane
realization.  After briefly reviewing some of the results of \refs{\OoguriBG\FrancoHT\BenaRG-\GiveonFK},
we calculate the leading terms in the low energy effective action of the model
and show that the gravitational interaction of the $D$-branes with an $NS5$-brane gives rise
to D-terms that play an important role in the existence of a metastable SUSY breaking vacuum.
We discuss the region in parameter space in which these D-terms give the leading contribution
to the potential of the pseudo-moduli, and the one in which this potential
is dominated by the one loop contribution computed in \IntriligatorDD.

In section 3 we turn to the generalized ISS model of \refs{\GiveonEF\GiveonEW-\EssigKZ}.
We exhibit all the metastable vacua described in \GiveonEW\ in the effective field theory and
discuss their fate as the parameters of the brane system are varied towards the renormalizable
field theory regime. As expected, many of the metastable states disappear in the process. In
section 4 we summarize our results and comment on them. Some technical details 
are described in the appendix.

\newsec{D-term supersymmetry breaking from branes}

In this section we discuss the string theory realization of the ISS model
\refs{\OoguriBG\FrancoHT\BenaRG-\GiveonFK}. As mentioned in the introduction, in string theory
the ISS pseudo-moduli are stabilized primarily by classical gravitational effects. We show that
in the low energy theory these effects give rise to non-trivial D-terms, which together with the
superpotential lead to a metastable SUSY breaking state.

\subsec{ISS from branes}

We start by decomposing the $9+1$ dimensional spacetime of type IIA
string theory as follows:
\eqn\decomp{\IR^{9,1}=\IR^{3,1}\times \IC_v\times \IC_w\times\IR_y\times\IR_{x^7}~,}
with
\eqn\nsns{v=x^4+ix^5~,\qquad w=x^8+ix^9~,\qquad y=x^6~.}
The brane configuration we consider is depicted in figure 1. All branes are extended in the
$\IR^{3,1}$ labeled by $(x^0, x^1,x^2,x^3)$. The $NS5$-branes denoted by $NS$ and
$NS'$ are further extended in $v$ and $w$, respectively. The $D6$-branes are extended
in $w$, $x^7$, while the $D4$-branes are stretched between other branes as indicated
in the figure.

\ifig\loc{The ISS brane configuration.}
{\epsfxsize5.0in\epsfbox{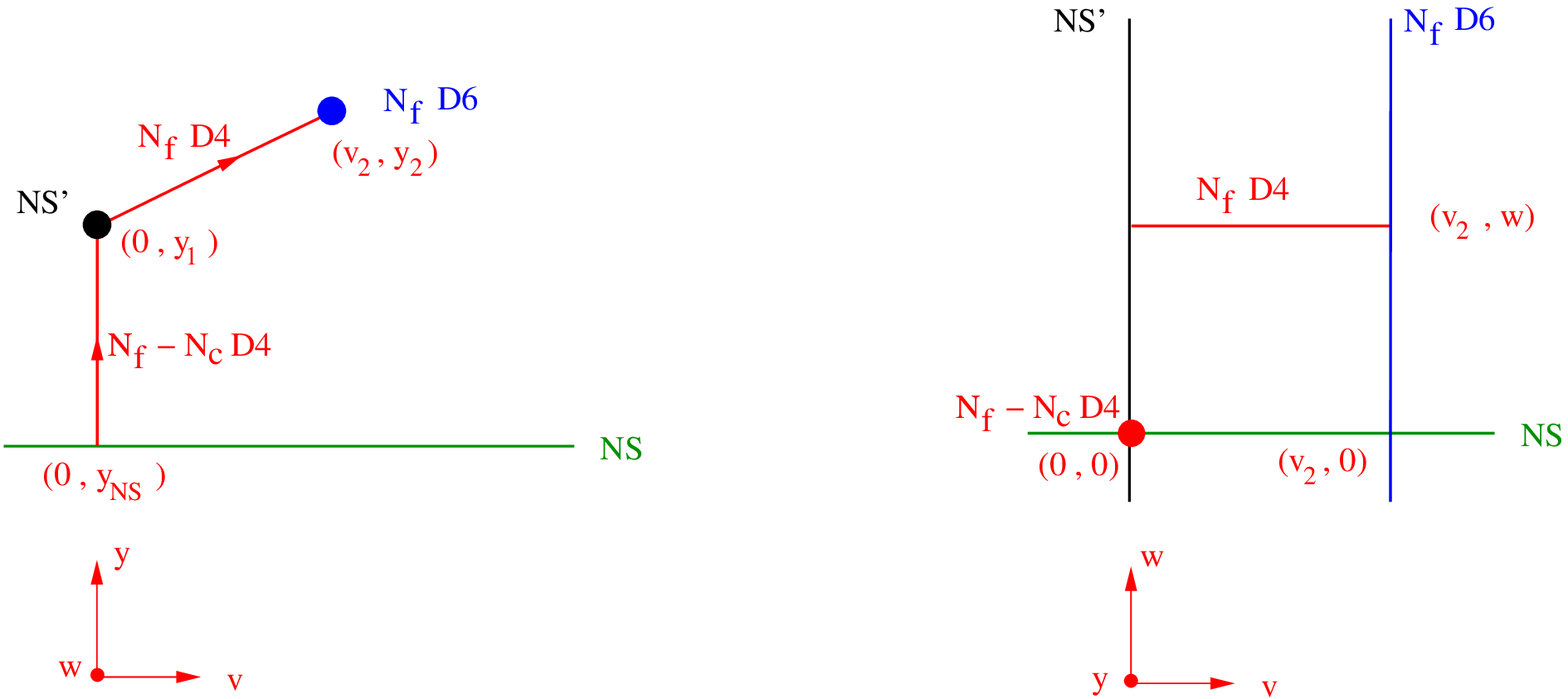}}

At low energies, the brane configuration of figure 1 reduces to an $N=1$ supersymmetric
gauge theory with gauge group $U(N_f-N_c)$, $N_f$ flavors of fundamentals $q^i,\tilde{q}_i$,
and gauge singlets $\Phi^{i}_{j}$, $i,j=1,\cdots, N_f$. The chiral superfields $q$,
$\tilde q$ and $\Phi$ have canonical K\"ahler potential,
\eqn\cankah{K=q^\dagger q+\tilde q^\dagger\tilde q+\Phi^\dagger\Phi~,}
and superpotential
\eqn\wmaggg{W_{\rm mag}=h  q^j\Phi^i_j \tilde q_i- h \mu^2 \Phi_i^i~.}
This theory is the Seiberg dual of $N=1$ SQCD with gauge group $U(N_c)$, and $N_f$
flavors $Q_i,\tilde Q^i$,  whose mass is proportional to $\mu^2$. It has been used to study
metastable SUSY breaking in  \IntriligatorDD.

The parameters of the low energy theory are given in terms of the underlying string theory
ones by
\eqn\hmug{g_{\rm mag}^2={4\pi^2g_sl_s\over y_1-y_{NS}}~, \qquad h^2={8\pi^2g_sl_s\over y_2-y_1}~,
\qquad \mu^2={v_2\over 16\pi^3 g_s l_s^3}~.}
Here and below we make a choice of phase such that $h$ and $\mu$ are real and positive.

The geometric description is reliable when the distances between the various branes are
large (relative to $l_s$), and the string coupling $g_s$ is small. In this regime, the
magnetic gauge coupling $g_{\rm mag}$ and Yukawa coupling $h$ are small. The mass parameter
$\mu$ is typically above the string scale. If $\mu$ is sufficiently small, the low
energy dynamics of the branes is well described by magnetic SQCD. In general, the low energy
effective Lagrangian receives contributions from other sources.

The brane configuration of figure 1 is unstable to reconnection of the  $N_f-N_c$ color $D4$-branes
with flavor $D4$-branes, leading to that of figure 2. The resulting configuration is marginally stable.
It contains an $N_c\times N_c$ matrix $X$ of massless fields describing the positions in the $w$-plane
of the $D4$-branes stretched between the $NS'$-brane and the $D6$-branes. The potential for
$X$ is flat since, as is clear from figure 2, the energy of the branes is independent of $w$.

\ifig\loc{The marginally stable brane configuration.}
{\epsfxsize5.0in\epsfbox{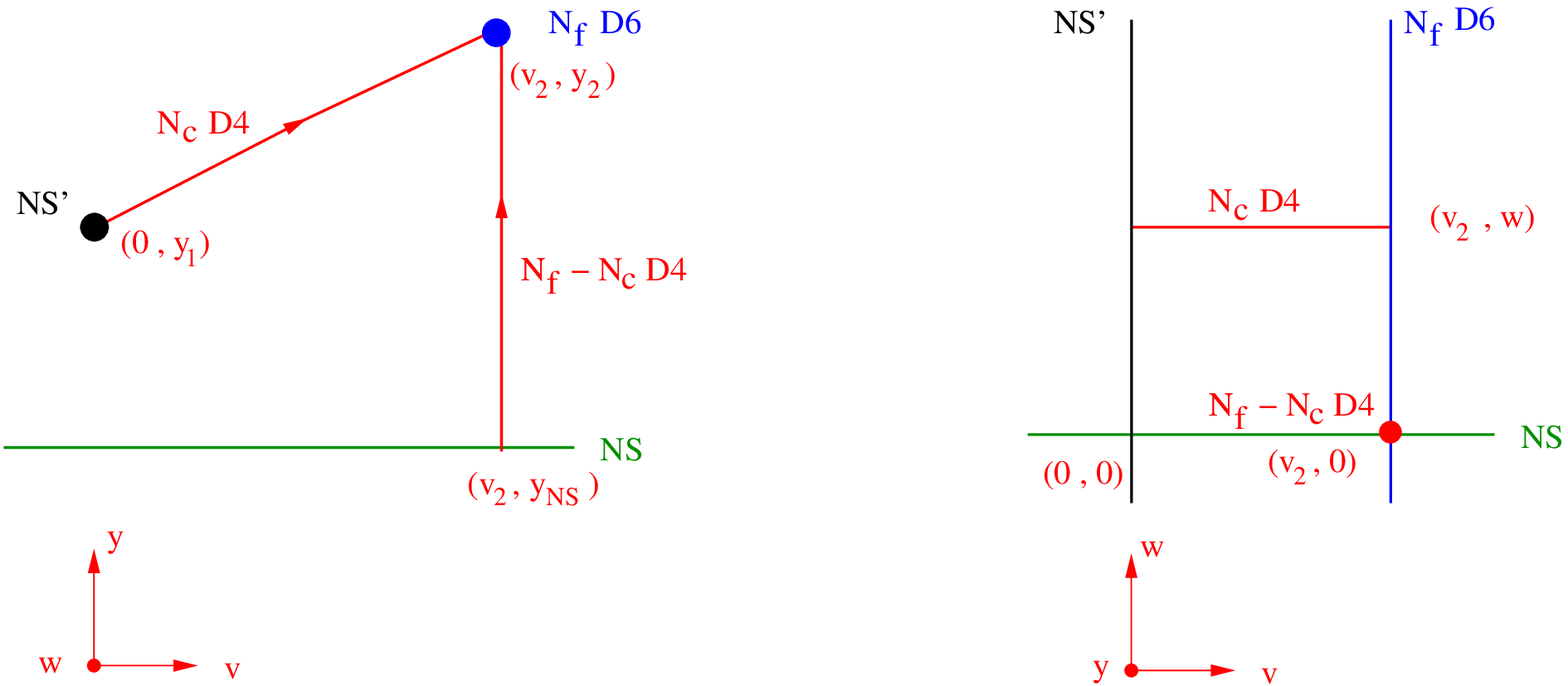}}

The above discussion has a simple analog in the effective field theory. The K\"ahler potential \cankah\
and superpotential \wmaggg\ give rise to the bosonic potential\foot{Following standard notation,
we denote the superfields and their bottom components by the same letter.}
\eqn\vveeff{V_0=h^2\left(|\tilde q q-\mu^2I_{N_f}|^2+|q\Phi|^2+|\Phi\tilde q|^2\right)~,}
where $I_{N_f}$ is a rank $N_f$ identity matrix.
The configuration of figure 1 describes the origin of field space (as is clear from the
fact that both the $U(N_f-N_c)$  gauge symmetry and the global $U(N_f)$ symmetry are
unbroken in it). Expanding \vveeff\ around this point we see that the magnetic quarks $q$,
$\tilde q$ have a tachyonic mass term. Thus, they get an expectation value,
which is the field theory analog of the brane reconnection process described above.

As is familiar from O'Raifeartaigh-type models, the potential \vveeff\ cannot be set to
zero due to the fact that the rank of $\tilde qq$, $N_f-N_c$, is smaller than that of the
identity matrix $I_{N_f}$. Hence, supersymmetry is broken. The minimum of the potential
corresponds (up to global symmetries) to
\eqn\mmqq{\tilde qq=\left(\matrix{ \mu^2 I_{N_f-N_c} & 0 \cr 0 & 0}\right)~,\qquad
\Phi=\left(\matrix{ 0 & 0 \cr 0 & X}\right)~.
}
$X$ is the $N_c\times N_c$ matrix field described in terms of the brane system above. It has a
classically flat potential, and thus parameterizes a pseudo-moduli space of non-supersymmetric vacua.

Since in the leading approximation the potential of  $X$ is
flat, we need to consider corrections. The nature of these corrections is different in
different regions of the parameter space of brane configurations. For $|v_2|\ll g_s l_s$,
the analysis of ISS is valid and the leading contribution to the potential on
pseudo-moduli space comes from one loop effects in the WZ model
\cankah, \wmaggg. The resulting Coleman-Weinberg potential behaves near the
origin like \IntriligatorDD
\eqn\voneloop{V_{1}={\ln4-1\over8\pi^2}(N_f-N_c)|h^2\mu|^2{\rm Tr X^\dagger X}+\cdots}
and gives a mass of order $|h^2\mu|$ to the pseudo-moduli.

For $v_2$ that remains finite in the limit $g_s\to 0$, the field theory potential \voneloop\ is a
subleading effect. The dominant contribution in this limit comes  from the gravitational attraction
of the $N_c$ $D4$-branes in figure 2 to the $NS$-brane  \GiveonFK. Our task in the rest of this
section is to understand  this gravitational effect in the low energy effective theory
of the light fields $q$, $\tilde q$ and $\Phi$.

\subsec{D-terms from branes}

It turns out to be useful to first consider the brane configuration of figures 1, 2 in the special
case $v_2=0$, for which $\mu=0$ (see \hmug) and supersymmetry, gauge and global symmetries
are unbroken. The WZ model \cankah, \wmaggg\ has in this case a moduli space of
supersymmetric vacua labeled by $\langle\Phi\rangle$, with $q=\tilde q=0$, along which the F-term
potential \vveeff\ vanishes.
In the brane construction, this moduli space corresponds to placing the flavor $D4$-branes at
arbitrary points in the $w$-plane.\foot{This gives $N_f$ of the $N_f^2$ moduli seen in the low energy
field theory. See \eg\ \GiveonSR\ for a description of the full moduli space.} It is easy to check that
the brane configuration preserves $N=1$ supersymmetry everywhere in moduli space, in agreement
with the field theory analysis.

In the above discussion, the brane configuration of figures 1, 2 (with $v_2=0$) is thought of as living
in flat spacetime. This is a good approximation when the distances between the branes are large,
but for our purposes it is important to include the leading corrections to this picture. Those are
due to the fact that the flavor $D4$-branes actually live in the geometry of the $NS$-brane.\foot{The
$NS'$ and $D6$-branes do not contribute to the discussion below since they are extended in $w$.}

To analyze the effects of this geometry on the moduli space, we consider the Dirac-Born-Infeld (DBI)
action for a $D4$-brane stretched in $y$ between the $NS'$ and $D6$-branes
 \eqn\DBI{\SS = -T_4 \int d^4 x \int_{y_1}^{y_2}dy e^{-\varphi} \sqrt{-{\rm det}P(G+B)_{ab}}~,}
where $T_4/g_s=1/(2\pi)^4l_s^5g_s$ is the tension of the $D4$-brane, $\varphi$ is the dilaton, and
$P(G+B)$ is the pullback of the spacetime metric and $B$-field to the worldvolume of the brane. The
$D4$-brane lives in the fivebrane geometry \CallanAT,
\eqn\chs{\eqalign{ &ds^2=dx_\mu dx^\mu+dv d\vbar + H(x^n) [ dy^2 + (dx^7)^2+dw d\wbar]~,\cr
&e^{2(\varphi-\varphi_0)}=H(x^n)~,\cr
&H_{mnp}=-\e_{mnp}^{~~~~~q}\partial_q\varphi~.\cr }}
Here $\mu=0,1,2,3$, while $m,n,p=6,7,8,9$ run over the directions transverse to the $NS$-brane.
$H_{mnp}$ is the field strength of the Neveu-Schwarz $B$ field sourced by the fivebrane;
$g_s=e^{\varphi_0}$ is the string coupling far from the fivebrane. The harmonic function $H$ is
given by
\eqn\hcoin{H(r)=1+{l_s^2\over r^2}~,}
with $r^2=x_mx^m=(y-y_{NS})^2+(x^7)^2+|w|^2$. The background \chs\ is reliable when $r\gg l_s$,
and we will assume this throughout our discussion.

In order to study the effects of the fivebrane background on the dynamics of the moduli, we take
the position of the $D4$-brane, $w$, to be a function of $x^\mu$, and plug it into the DBI
action \DBI. The resulting four dimensional effective Lagrangian is
 \eqn\DBIN{\CL = -{T_4\over g_s} \int_{y_1}^{y_2} dy  \sqrt{1+H(r)|\partial_\mu w|^2
 -{1\over8}H^2(\partial_\mu w\partial_\nu\bar w-\partial_\mu\bar w\partial_\nu w)^2}~.}
Taking $w(x^\mu)$ to be constant gives
\eqn\wconst{\CL=-{T_4\over g_s}\Delta y~,}
the (negative of) the energy density of a BPS $D4$-brane of length $\Delta y=y_2-y_1$ in flat space.
As expected from supersymmetry, the fivebrane background does not lift the moduli space labeled by $w$,
and does not modify the energy of the brane.

The next term in the expansion of the square root in \DBIN\ gives the kinetic term of $w$,
\eqn\wkin{\CL_2=-{T_4\over2g_s}\int_{y_1}^{y_2} dy  H(r)|\partial_\mu w|^2~.}
In general, the integral in \wkin\ depends non-trivially on all the parameters. It simplifies
in the limit where the length of the $D4$-brane, $\Delta y$, is much smaller than its distance
to the $NS$-brane, \ie\ $l_s\ll\Delta y\ll |y_{NS}-y_i|$. One can further choose the origin of
$y$ such that the $NS'$ and $D6$-branes are located near the origin, with the $NS$-brane far
away from them, \ie\ $l_s\ll |y_1|,|y_2|\ll |y_{NS}|$. In that case, one has
\eqn\intkin{\int_{y_1}^{y_2} {dy\over (y-y_{NS})^2+|w|^2}\simeq {\Delta y\over y_{NS}^2+|w|^2}~,}
so
\eqn\wkinnn{\CL_2=-{T_4\over2g_s}\Delta y  H(|w|)|\partial_\mu w|^2~,}
where
\eqn\hyns{H(|w|)=1+{l_s^2\over y_{NS}^2+|w|^2}~.}
In this limit the $D4$-brane behaves like a local probe of the geometry of the fivebrane.

Equations \intkin\ -- \hyns\ capture the full dependence of the kinetic term \wkinnn\ on the
parameter $w/y_{NS}$ but neglect corrections that vanish in the limit $y_i/y_{NS}\to 0$.
As $y_{NS}\to-\infty$, $H\to 1$, and the Lagrangian \wkinnn\ takes a canonical form
in terms of the field
\eqn\cannorm{\Phi=w\sqrt{T_4\Delta y\over 2g_s}~.}
This relation  can be alternatively written as
\eqn\hphiw{h\Phi={w\over2\pi l_s^2}~,}
which is obtained by comparing the mass  of $q$, $\tilde q$ at a point $\langle\Phi\rangle$ in
moduli space to the energy of a fundamental string of length $w$.

For finite $y_{NS}$, the effect of the fivebrane geometry is to generate a
non-trivial K\"ahler potential on the moduli space labeled by $\Phi$,
\eqn\KahlerNoAngle{\partial_\Phi\partial_{\Phibar}\KK\equiv\KK_{\Phi\Phibar}
= H(|\Phi|)=1+{l_s^2\over y_{NS}^2+
|w(\Phi)|^2}=1+{l_s^2\over y_{NS}^2}\left(1+{|\Phi|^2\over\Lambda^2}\right)^{-1}~,}
with
\eqn\llaamm{\Lambda^2={T_4 y_{NS}^2\Delta y\over 2g_s}=\left(y_{NS}\over 2\pi l_s^2 h\right)^2~.}
Near the origin, the K\"ahler potential \KahlerNoAngle\ behaves as
\eqn\kahsmall{\KK_{\Phi\Phibar}\simeq 1-{|\Phi|^2\over\tilde\Lambda^2}\left[
1+O\left(|\Phi|^2\over\Lambda^2\right)\right]~,}
where
\eqn\lamtil{\tilde\Lambda={y_{NS}\over l_s}\Lambda={y_{NS}^2\over 2\pi l_s^3 h}~,}
and we neglected subleading terms for $l_s/y_{NS}\ll1$. 

In the low energy theory, the
non-canonical K\"ahler potential \kahsmall\ corresponds to including in the effective
Lagrangian an infinite series of higher dimension operators. Such operators are typically
suppressed by a UV scale. In our case, two such scales appear, $\Lambda$ \llaamm\ and
$\tilde\Lambda$ \lamtil. For $y_{NS}\gg l_s$, the regime in which the geometric description
is valid, they are widely separated, $\tilde\Lambda\gg\Lambda$. All other dimensionful
parameters in the model can be expressed in terms of $h$, $\mu$, $\Lambda$, $\tilde\Lambda$.
E.g. the string scale is given by
\eqn\stscale{2\pi l_s={h\tilde\Lambda\over(h\Lambda)^2}~;}
$y_{NS}$ can be computed by plugging \stscale\ into \lamtil.

So far we took $\Phi$ to be a single chiral superfield, or in the brane language restricted to
the case $N_f=1$. In general, $\Phi$ is an $N_f\times N_f$ matrix and the kinetic term \wkin\
can be written as (the bosonic part of)
\eqn\lkinfin{\CL_2=\int d^4\theta {\rm Tr} \KK(\Phi,\Phi^\dagger)~,}
with the K\"ahler potential given by \KahlerNoAngle.

In the discussion above we focused on the bosonic terms of the Lagrangian \lkinfin. Terms involving
the fermions are related to them by supersymmetry, which is preserved by the DBI action.  Therefore,
they are in principle guaranteed to agree with \lkinfin. In appendix A we verify that this is indeed the case.

The K\"ahler potential \KahlerNoAngle\ is not the only effect of the fivebrane on $\Phi$.
Expanding the square root \DBIN\ one finds derivative interactions associated
with higher order D-terms. To illustrate this, consider the next (quartic) term in the expansion of \DBIN,
\eqn\lfour{\eqalign{\CL_4=&{T_4\over16g_s}\int_{y_1}^{y_2}dy H^2(r)\left[2|\partial_\mu w|^4
+(\partial_\mu w\partial_\nu\bar w-\partial_\mu\bar w\partial_\nu w)^2\right]\cr
=&{T_4\over8g_s}\int_{y_1}^{y_2}dy H^2(r)|\partial_\mu w\partial^\mu w|^2
={g_s\over 2T_4\Delta y}H^2(|\Phi|)|\partial_\mu\Phi\partial^\mu\Phi|^2~.\cr}}
In the last equality we used the approximation discussed after eq. \wkin, and \cannorm.

In the low energy theory of the field $\Phi$, \lfour\ is due to the following D-term:
\eqn\DFour{\CL_4= \int d^4\theta G(\Phi,\Phibar)\epsilon^{\alpha\beta}\epsilon^{\dot\alpha\dot\beta}
\DD_\alpha\Phi \DD_\beta\Phi\bar\DD_{\dot\alpha}\Phibar\bar\DD_{\dot\beta}\Phibar~.}
Using the standard conventions for the expansion of a chiral superfield (see \eg\ \WessCP),
and performing the $\theta$ integrals, the bosonic terms in \DFour\ take the form
\eqn\Ggg{\CL_4=16 G(\Phi,\Phibar) \left(|F|^4 - 2|F|^2 |\del_\mu \Phi|^2 + |\partial_\mu\Phi\partial^\mu\Phi|^2\right).~}
Setting $F=0$ by using its equation of motion and comparing the four-derivative term in \Ggg\ with
\lfour\ we see that
\eqn\GTwo{G(\Phi,\Phibar) = {g_s\over32 T_4\Delta y}H^2(|\Phi|)~.}
Higher order terms in the expansion of the square root \DBIN\ give an infinite series of higher
order D-terms, which can be calculated in the same way.

\subsec{Supersymmetry breaking at non-zero $\mu$}

We now turn to the general situation in figure 2, with $v_2\not=0$, for which
supersymmetry is broken. As is clear from the figure, the origin of the breaking is twofold.
First, the $N_c$ $D4$-branes which support the (pseudo-)moduli $X$ \mmqq\ are not mutually
BPS with the $N_f-N_c$ remaining $D4$-branes. As a consequence, the spectrum of open strings
connecting the two stacks of branes  is non-supersymmetric. These open strings are massive,
and can be integrated out. Since they must appear in pairs in intermediate states, they influence
the low energy dynamics of $X$ only via string loops. We are interested in effects that survive in
the classical limit $g_s\to 0$; hence, we will ignore them.

The second source of supersymmetry breaking in figure 2 is the $NS$-brane, which is also
not mutually BPS with the $N_c$ $D4$-branes. This gives rise to supersymmetry breaking effects
that survive in the classical limit, and are sensitive to the distance between the fourbranes
and fivebrane, $|y_{NS}|$, as well as to the position of the $D6$-branes in the $v$-plane, $v_2$.
Our goal in this subsection is to analyze these effects in the low energy theory.

The $N_c$ $D4$-branes in figure 2 are stretched along the line
\eqn\angledf{v=a(y-y_1)~.}
The slope $a$ is given by
\eqn\slopepar{a={v_2\over\Delta y} = 2\pi (l_s h \mu)^2~,}
where we used \hmug\ to relate $a$ to the field theory parameters.
Plugging \angledf\ into the DBI action \DBIN\ leads\foot{As in the previous subsection, for the purpose
of this discussion we can consider each of the $N_c$ $D4$-branes individually. Thus, we set $N_c=1$
below; the generalization to larger $N_c$ is straightforward.} to the following effective Lagrangian for $w$:
 \eqn\DBIangle{\CL = -{T_4\Delta y\over g_s}
  \sqrt{1+{a^2\over H}}\sqrt{1+H|\partial_\mu w|^2+O(|\partial w|^4)}~.}
Setting $w$ to a constant, we see that the energy of the $D4$-brane is no longer
independent of $w$,
\eqn\endfour{V={T_4\Delta y\over g_s} \left(\sqrt{1+{a^2\over H}}-1\right)~,}
where we have subtracted from \DBIangle\  the energy of the supersymmetric configuration
(the one with $a=0$), as is standard in supersymmetric field theory \WessCP.

We will find it convenient to expand in the slope $a$ \slopepar. To leading order,
\endfour\ takes the form
\eqn\KahlerTwo{V =\frac{T_4 \Delta y}{2 g_s} \frac{a^2}{H} + \OO(a^4)~.}
In the low energy theory of $\Phi$, this must be due to a non-zero superpotential.
In general one has \WessCP:
\eqn\Sup{V={|\partial_\Phi\WW|^2 \KK^{\Phi \Phibar}}~.}
Plugging in the potential \KahlerTwo\ and K\"ahler potential \KahlerNoAngle\ one finds:
\eqn\SupTwo{ |\partial_\Phi\WW|^2 = {T_4 \Delta y\over2g_s}a^2=|h\mu^2|^2~.}
In the last equality we used the explicit forms of $h$, $\mu$, \hmug. We see that
the DBI calculation is compatible with the superpotential $W=-h\mu^2\Phi$ \wmaggg,
and K\"ahler potential \KahlerNoAngle.

Expanding the potential \Sup\ around the origin of pseudo-moduli space, we find
\eqn\massform{V=|h\mu^2|^2+M_X^2|X|^2+O(|X|^4)~,}
with
\eqn\massphi{M_X=a{l_s\over y_{NS}^2}={h\mu^2\over\tilde\Lambda}~.}
The first equality shows that the pseudo-modulus $X$ \mmqq\ develops a mass linear
in $a$. The second expresses this mass in terms of the parameters appearing in the low
energy Lagrangian, where it is a consequence of the non-zero superpotential \SupTwo\
{\it and} non-canonical K\"ahler potential \kahsmall. In particular, if we send
$y_{NS}\to-\infty$, or in field theory language $\tilde\Lambda\to\infty$, the mass
\massphi\ goes to zero, and supersymmetry is restored, despite the fact that the linear
superpotential \wmaggg\ remains non-trivial in this limit.

This is due to the familiar fact that a chiral superfield with a linear superpotential
and canonical K\"ahler potential describes a free massless boson and fermion and does
not really break supersymmetry. More formally, one can in that case shift $\Phi$ by its
non-zero F-term and recover a Lagrangian with vanishing superpotential. When the
K\"ahler potential is non-trivial, this cannot be done.

From the brane perspective this is natural as well, since when the $NS$-brane is absent (and
ignoring the $N_f-N_c$ $D4$-branes in figure 2, as discussed above), the brane system is
supersymmetric, and the only difference with respect to the system with $a=0$ is that the
length of the $N_c$ $D4$-branes, and thus their energy, is $a$-dependent. This $a$ dependence
leads to the linear superpotential of the low energy effective field theory.

Note also that, as mentioned above, the potential on pseudo-moduli space due to the non-trivial
K\"ahler potential \massform, \massphi, is much larger than that due to the ISS one loop potential,
\voneloop. Indeed, the former remains finite as $g_s\to 0$ with finite $a$, while the latter goes
to zero like $g_s$. This is due to the fact that the  K\"ahler potential is a consequence of classical
gravitational interactions, while \voneloop\ is a one loop effect. From the point of view of the low
energy effective field theory, the above discussion is valid when the UV scale $\tilde\Lambda$ is much
smaller than $\mu/h$ (neglecting numerical factors).

To recapitulate, the kinetic terms for the light bosons and fermions at order $a^0$, as well as the
potential for the scalars at order $a^2$ \KahlerTwo\ are compatible with the K\"ahler potential
\KahlerNoAngle, and superpotential \wmaggg, which go like $a^0$ and $a$, respectively. All these
terms have been computed to leading order in $a$. We next comment on the structure of subleading
corrections.

One can think of the perturbative expansion in $a$ as follows. The K\"ahler potential and superpotential
discussed above give a mass proportional to $a$, \massphi, to the pseudo-moduli. This mass defines a
natural momentum scale in the low energy effective theory. Therefore, when expanding in powers of $a$,
derivatives $\partial_\mu$ scale like $a$, super-derivatives $\DD_\alpha$ like $\sqrt a$, etc.
The bottom component of the chiral superfield  $\Phi$ scales like $a^0$, while the fermion $\psi$ scales
like $\sqrt a$ due to the factor of $\theta$ in front of it in the component expansion of the chiral superfield.

According to these rules the K\"ahler potential and superpotential terms computed above scale like $a^2$.
The D-term \DFour\ computed in the previous subsection scales like $a^4$. Other possible terms that scale
like $a^4$ include a correction of order $a^2$ to the K\"ahler potential \KahlerNoAngle, and a contribution
proportional to $a^3$ to the superpotential  \wmaggg, \SupTwo.

In order to calculate these terms one can proceed as follows. From \DBIangle\ we see that the kinetic
term for $\Phi$ receives a correction at order $a^2$:
\eqn\ltwocorr{\delta\LL_2=-\half a^2|\partial_\mu\Phi|^2~.}
In the low energy effective action there are two possible contributions to \ltwocorr. One is from the
D-term \DFour. Plugging the (order $a$) F-term,
\eqn\ffttee{|F|^2={|\partial_\Phi W|^2\over |\KK_{\Phi\Phibar}|^2}={a^2T_4\Delta y\over2g_s H^2}~,}
into \Ggg, we find that the D-term \DFour\ exactly accounts for the correction \ltwocorr. Since the
only other contribution to the kinetic term at this order is from the K\"ahler potential,
we conclude that \KahlerNoAngle\ is in fact not corrected at order $a^2$.

We next turn to the superpotential. To leading order in $a$, it goes like
\eqn\wwss{W=-a\sqrt{T_4\Delta y\over 2g_s}\Phi~.}
Since $\Phi$ is charged under a $U(1)$ global symmetry that corresponds geometrically
to rotations of the $w$-plane, any higher order corrections to \wwss\ must be proportional
to $\Phi$ as well. At the order under consideration, the only possible correction
goes like $\delta W\sim a^3\Phi$. To calculate it, we expand \endfour\ to order $a^4$:
\eqn\KahlerFour{V =\frac{T_4 \Delta y}{2 g_s} \frac{a^2}{H} - \frac{T_4 \Delta y}{8 g_s} \frac{a^4}{H^2}+
\OO(a^6)~.}
Since the K\"ahler potential is not corrected at order $a^2$, and thus does not contribute to the
$a^4$ term in \KahlerFour, there are again two possible contributions to this term. One comes from
the D-term \DFour.  Plugging  \ffttee\ into \Ggg\ gives
\eqn\potcont{V_4=-16G|F|^4=-\frac{T_4 \Delta y}{8 g_s} \frac{a^4}{H^2}~.}
Comparing to \KahlerFour\ we see that this agrees with the DBI calculation.
Therefore, we conclude that the superpotential \wwss\ does not receive corrections
at order $a^3$.

To summarize, we find that to order $a^4$ in the expansion of the effective Lagrangian described
above, the K\"ahler potential and superpotential are given by \KahlerNoAngle\ and \wwss, respectively.
Corrections to these potentials that are down by $a^2$ from the leading contributions vanish. The non-zero
corrections to the DBI results at this order are due to the higher derivative D-term \DFour.

It is natural to expect that this pattern persists to higher orders in $a$. Indeed, the K\"ahler potential
and higher D-terms should not depend on the orientation of the $D4$-brane (labeled by $a$ \angledf) in the
limit $\Delta y\ll y_{NS}$, in which the fourbrane can be considered as a local probe of the geometry. The
form of the superpotential is determined by the $SO(2)_v\times SO(2)_w$ R-symmetry (corresponding to rotations in the $v$ and $w$ planes, respectively) to be linear in $v_2$ and $w$, or equivalently in 
$\mu^2$ and $\Phi$.

One can use the DBI action \DBIangle\ to calculate the mass of the pseudo-moduli \massphi\ to all
orders in $a$. A short calculation gives
\eqn\shortcalc{M_X={a\over\sqrt{1+a^2}}{l_s\over y_{NS}^2}~.}
As discussed above, from the point of view of the low energy effective action,
this mass receives contributions from the whole infinite tower of D-terms.

For $a$ that remains finite in the limit $g_s\to 0$, the contribution to the mass of the pseudo-moduli 
\shortcalc\ of the non-canonical K\"ahler potential, \massphi, and other D-terms, is much larger than that
computed in the low energy field theory in \IntriligatorDD, \voneloop. On the other hand, if one fixes
all the parameters other than $v_2$ and sends $v_2\to 0$, the ISS mass \voneloop\ eventually becomes larger,
since it goes like $\sqrt{v_2}$, while \massphi\ goes like $v_2$. Formally, the two masses become comparable
when
\eqn\compmass{v_2\sim {g_s y_{NS}^4\over l_s^3}~.}
However, in this regime both calculations are unreliable. The DBI calculation fails because $v_2$ \compmass\
and thus the mass \massphi\ are of order $g_s$, and one has to include string loop effects in the discussion.
The field theory result \voneloop\ is not valid since \compmass\ corresponds to
\eqn\mumu{\mu\sim m_s\left(y_{NS}\over l_s\right)^2~,}
which is much larger than $m_s$. As mentioned above, the field theory analysis is reliable
for $\mu\ll m_s$, which corresponds to $v_2\ll g_s l_s$.

\newsec{R-symmetry breaking metastable vacua}

In this section we turn to the generalized ISS model of \refs{\GiveonEF\GiveonEW-\EssigKZ}.
We use the results of the previous section to describe the R-symmetry breaking metastable
vacua of \GiveonEW\ in the low energy effective field theory, and discuss their fate as the
parameters of the brane system are varied.

\subsec{Deformed ISS from branes}

The brane system we consider is depicted in figure 3. It is obtained from that of figure 2
by a rotation of the $D6$-branes by an angle $\theta$ in $(v,w)$. In the low energy theory,
the rotation of the sixbranes corresponds \GiveonEW\ to adding a $\Phi^2$ term to the
superpotential \wmaggg:
\eqn\wdef{W=hq\Phi\tilde q -h\mu^2\Tr\Phi+\half h^2\mu_\phi\Tr\Phi^2~.}
The parameters $h$, $\mu$ are again given by \hmug, while
\eqn\muphi{ \mu_\phi = {\tan{\theta} \over 8\pi^2 g_s l_s }~.}
\ifig\loc{The deformed ISS brane configuration.}
{\epsfxsize5.0In\epsfbox{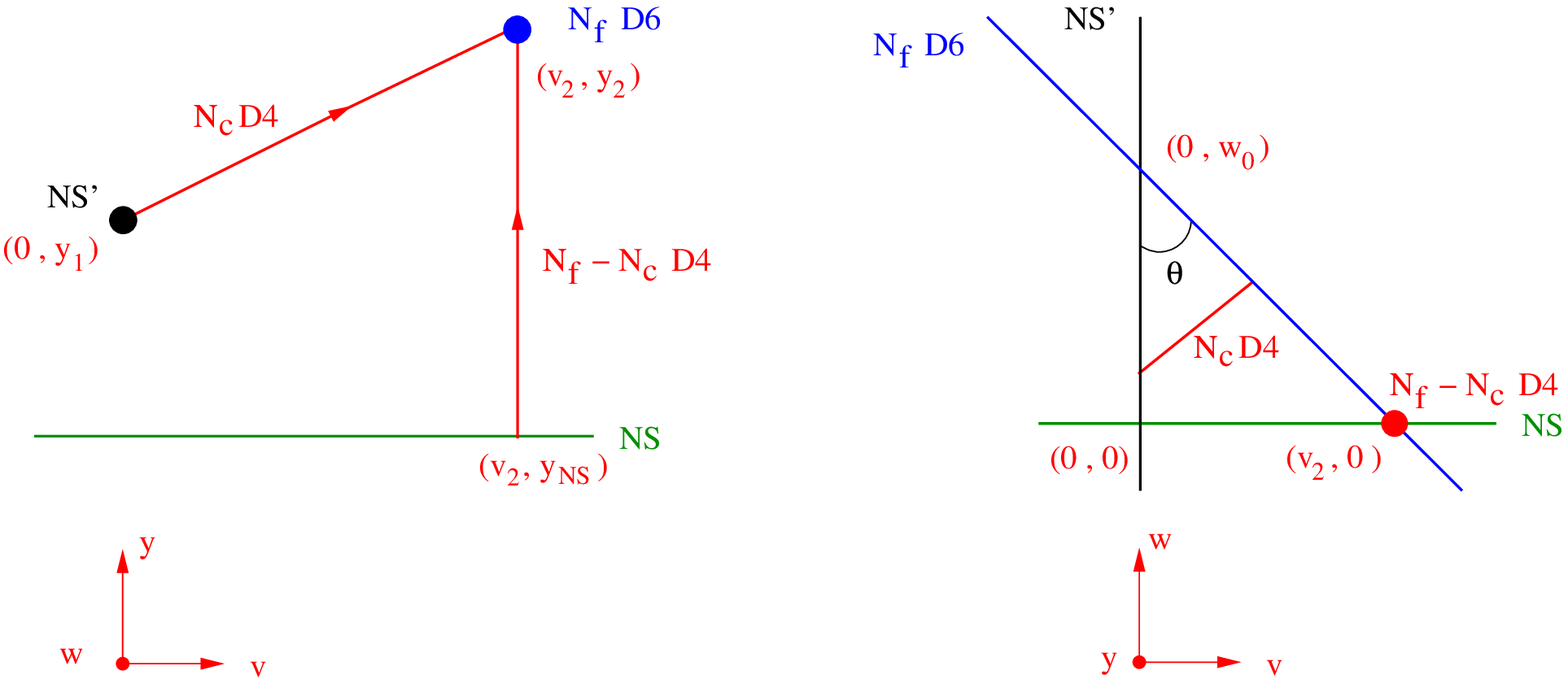}}

The orientation of the $D6$-branes leads to an important difference
in the dynamics of the configuration of figure 3 compared to that of
figure 2. While the latter has (to leading order) a pseudo-moduli
space of supersymmetry breaking vacua labeled by the positions of the
$N_c$ $D4$-branes in $w$, the former has a supersymmetric vacuum,
in which the $D4$-branes are located at the point $(v,w)=(0,w_0)$, with
\eqn\intbranes{|w_0|=|v_2|\cot\theta~.}
At that point, the projections of the $NS'$ and $D6$-branes on the $(v,w)$
plane intersect, and the $D4$-branes stretch between them along the $y$ direction.
In fact, the brane system has many supersymmetric vacua, which are described
and compared to field theory in \GiveonEW. From the perspective of the WZ
model with superpotential \wdef, these vacua are obtained by analyzing
the zeroes of the bosonic potential corresponding to \wdef,
\eqn\vveeffdef{V_{0}=h^2\left(|\tilde q q-\mu^2I_{N_f}+h\mu_\phi\Phi|^2
+|q\Phi|^2+|\Phi\tilde q|^2\right)~.}

\noindent
To study metastable vacua in the brane system, one needs to add to the
above discussion the effects of the $NS$-brane background discussed in the
previous section. This leads to a force attracting the $D4$-branes in figure
3 towards $w=0$, as before. This force counteracts the one coming from \vveeffdef,
which pushes the $D4$-branes away from the origin. The balance of the two effects
generically leads to the appearance of metastable supersymmetry breaking vacua
at $w<w_0$ \GiveonEW. Our main task in this section is to understand these vacua
in terms of the low energy field theory of the fields $q$, $\tilde q$, $\Phi$.

\subsec{ISS-type vacua}

As in section 2, in studying the metastable vacua of the brane configuration of figure 3
we will focus on one of the $N_c$ $D4$-branes stretched between the $NS'$ and $D6$-branes,
and neglect the remaining $N_f-N_c$ $D4$-branes, whose effects are down by $g_s$. We will
also continue to work in the regime $|y_i|\ll |y_{NS}|$, and neglect the curving of the
$D4$-brane in the fivebrane background (which is discussed in \refs{\GiveonFK,\GiveonEW}).

The $D4$-brane under consideration is displaced from the supersymmetric vacuum at
$(v,w)=(0,w_0)$, and stretches between the $NS'$ and $D6$-branes along a straight line in
$(v,y,w)$ space. This line can be parameterized by a variable $\lambda\in[0,1]$ as follows:
\eqn\DfourPar{ (v,y,w) = (\lambda (w_0 - w_1)\sin{\theta} \cos{\theta} ,
\lambda \Delta y + y_1, \lambda (w_0 - w_1) \sin^2{\theta} + w_1 )~. }
$\lambda=0, 1$ correspond to the endpoints of the $D4$-brane
lying on the $NS'$ and $D6$-branes, respectively.

To study the dynamics of the fourbrane  one can \eg\ take $w_1$ (the position of  the
endpoint of the $D4$-brane on the $NS'$-brane) to be a dynamical field $w_1(x^\mu)$,
and plug \DfourPar\ into the DBI action \DBI. This gives
\eqn\ISSDefDBI{{\cal S } = - T_4 \int d^4 x \int_0^1d\lambda
e^{-\varphi} \sqrt{g_{\lambda\lambda}}
 \sqrt{ -\det{( \eta_{\mu\nu} + H \partial_\mu w_1 \partial_\nu \bar{w}_1 )} } ~,}
where
\eqn\Inducedg{ e^{-(\varphi-\varphi_0)} \sqrt{g_{\lambda\lambda}} =
\left[ (\Delta y)^2 + |w_0 - w_1|^2 \sin^2{\theta} \left( \sin^2{\theta} +
\frac{\cos^2{\theta}}{H} \right) \right]^{1\over2}~,}
and
\eqn\ISSDefH{ H = 1 + \frac{l_{s}^2}{ y_{NS}^2 + |\lambda \sin^2{\theta} (w_0 - w_1) + w_1|^2 }~.}
The bosonic potential is given by (after subtracting the energy of the supersymmetric configuration)
\eqn\ISSDefVzero{V = {T_4\over g_s}\int_0^1 d\lambda
\left[ e^{-(\varphi-\varphi_0)} \sqrt{g_{\lambda \lambda}} - \Delta y \right]~.}
In general, the $D4$-brane \DfourPar\ has a finite extent in $w$, and the integrals
in \ISSDefDBI, \ISSDefVzero\ are non-trivial. The situation simplifies for small $\theta$,
where the extent of the fourbrane in $w$, $|w_0 - w_1| \sin^2{\theta}$ is small (of order
$\theta$) and the integrands of the above integrals can be taken to be approximately
independent of $\lambda$. In that limit, the $D4$-brane becomes a local probe of the
geometry of the fivebranes, as in section 2. To leading order in $a$ \slopepar\ and $\theta$,
one can describe its location in the $w$ plane by a complex scalar field, $w(x^\mu)$, whose
kinetic term is given by \wkinnn, or by the corresponding canonically normalized field $\Phi$
defined in \cannorm.

The bosonic potential \ISSDefVzero\ is given to leading order in $\theta$ by
\eqn\ISSDefVone{ V = \frac{T_4}{2 g_s \Delta y H} \theta^2 | w_0 - w |^2 =
\frac{1}{H} \left| v_2 \sqrt{ \frac{ T_4}{ 2 g_s \Delta y} } - \frac{ \theta}{\Delta y} X \right|^2 ~.}
In the second equality, $X$ is the field defined in \mmqq.
Using  \hmug\ and \muphi, \ISSDefVone\ may be written  as
\eqn\ISSDefV{ V_0 = \frac{1}{H} | h \mu^2 - h^2 \mu_\phi X |^2 ~.}
Comparing to \Sup\ we see that the potential \ISSDefV\ agrees with the one obtained
from the superpotential \wdef\ and K\"ahler potential \KahlerNoAngle. The potential
vanishes at
\eqn\susyvac{ X_{\rm susy}={\mu^2\over h\mu_\phi}~,}
the supersymmetric vacuum. For small $\mu_\phi$ this
vacuum is located far from the origin of field space.

Near the origin of field space, the potential \ISSDefV\ behaves as follows:\foot{Recall
that the parameters $h$, $\mu$, $\mu_\phi$ are taken to be real and positive.}
\eqn\nearorig{V_0\simeq h^2\left(\mu^4-2h\mu_\phi\mu^2X+
\mu^4{X^2\over\tilde\Lambda^2}+\cdots\right)~,}
where the $\cdots$ stand for terms that are negligible for small $\mu_\phi$.
In \nearorig\ we neglected the quadratic term in the expansion of the
classical potential \ISSDefV\ relative to the one that comes from the K\"ahler potential
\kahsmall. This is justified when
\eqn\negclass{h\tilde\Lambda\ll {\mu^2\over\mu_\phi}~,}
\ie\ the UV scale $\tilde\Lambda$ is well below the value of $X$ at the supersymmetric vacuum,
\susyvac.

The potential \nearorig\ has a local minimum at
\eqn\metmin{hX_{\rm min}={\mu_\phi\over\mu^2}(h\tilde\Lambda)^2~.}
In the regime \negclass\ this minimum is located well below $\tilde\Lambda$. Since in
obtaining \metmin\ we approximated the K\"ahler potential \KahlerNoAngle\ by \kahsmall,
it is valid when  $X_{\rm min}\ll\Lambda$. This implies the following hierarchy of scales:
\eqn\hierarchy{X_{\rm min}\ll\Lambda\ll\tilde\Lambda\ll\sqrt{\Lambda X_{\rm susy}}
\ll X_{\rm susy}~,}
where $X_{\rm susy}$ is given by \susyvac.

We see that the brane analysis of the R-symmetry breaking brane configuration
of figure 3 performed in \GiveonEW\ has a simple interpretation in the low energy effective
field theory. Combining the non-trivial K\"ahler potential found in section 2, \KahlerNoAngle,
with the superpotential corresponding to the deformed brane system, \wdef, leads to metastable
vacua at non-zero $\Phi$ \metmin, in which the two balance each other. As in section 2, the
corresponding WZ model \refs{\GiveonEF,\EssigKZ} exhibits similar vacua, but the stabilization
mechanism is different -- the role of the K\"ahler potential is now played by the one loop
potential. We will comment on the transition between the two regimes later.

\subsec{Tachyonic branches}

The brane system of figure 3 has a rich phase structure that was explored in \GiveonEW, where it was
found that  vacua are labeled by two integers $k$, $n$ (see figure 4). The $N_f-N_c-k$ $D4$-branes
stretched between the $NS$ and $NS'$-branes give rise to an unbroken $U(N_f-N_c-k)$ subgroup
of the magnetic gauge group. A $U(k)\times U(n)\times U(N_f-k-n)$ subgroup of the $U(N_f)$ flavor
group is unbroken as well.

The magnetic meson field $\Phi$ can be decomposed as
\eqn\metaformphi{h \Phi=\left(\matrix{0_k & 0 & 0\cr 0 & h \Phi_n & 0 \cr 0 & 0 &
{\mu^2 \over \mu_\phi}I_{N_f-k-n}\cr}\right)~.}
The $n\times n$ matrix $\Phi_n$ describes the position of the $n$ fourbranes that break supersymmetry in
figure 4. The system can relax to a supersymmetric state in one of two ways. The $n$ $D4$-branes can move
to $w=w_0$, \intbranes, where $h\langle\Phi_n\rangle={\mu^2\over\mu_\phi}I_n$; this leads to the
configuration of figure 4 with $n=0$. Alternatively, they can connect with some of the $N_f-N_c-k$ color
branes, effectively increasing the value of $k$. Of course, in general the system can relax by a
combination of the two processes.

\ifig\loc{The vacuum structure of the deformed brane configuration.}
{\epsfxsize5.0in\epsfbox{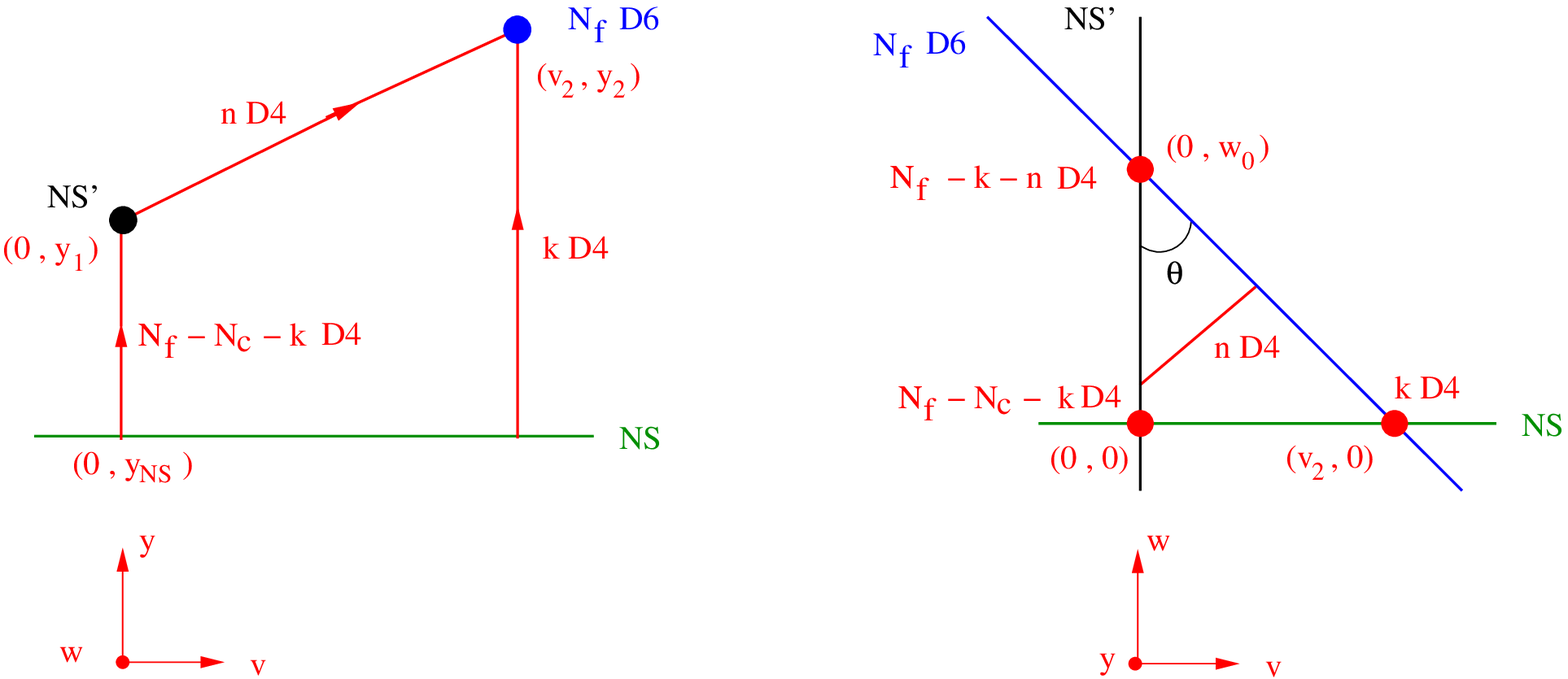}}

In addition to the supersymmetric vacua, the system has some non-supersymmetric metastable ones.
For $k=N_f-N_c$ they can be constructed using the results of the previous subsection. Here, we
would like to discuss the case $k<N_f-N_c$, where some new issues appear due to the presence of
the color branes.

The analysis of the effective potential for $\Phi_n$ that is due to the K\"ahler potential  \KahlerNoAngle\
and superpotential \wdef\ is identical to the one for $X$ in the previous subsection. It leads to a local
minimum of the potential at (see \metmin)
\eqn\phinmin{h\langle\Phi_n\rangle={\mu_\phi\over\mu^2}(h\tilde\Lambda)^2I_n~.}
However, in this case there is a further potential instability. In brane language it
corresponds to the reconnection process of some or all of the $n$ flavor branes with
color ones. It is described in string theory by the condensation of an open string
tachyon stretched between the two stacks of branes. The mass of this tachyon is given by
\PolchinskiRR
\eqn\masstach{\alpha'm^2=-{a\over2\pi}~.}
For it to be non-tachyonic, the endpoints of the color and flavor branes on the $NS'$-brane must
be a distance $w$ apart, with
\eqn\wapart{w>l_s \sqrt{2\pi a}~.}
Using \hphiw\ and \slopepar, \wapart\ implies that
\eqn\constphi{\langle\Phi_n\rangle  =x\mu I_n, \qquad {\rm with}\;\;   x>1 ~.}
Comparing to \phinmin\ we conclude that from the brane perspective,
the vacua with $k<N_f-N_c$ are metastable when
\eqn\concorr{ \tilde\Lambda>\tilde\Lambda_c=\sqrt{\mu\cdot {\mu^2\over h\mu_\phi}}=\sqrt{\mu X_{\rm susy}}~.}
Note that  \concorr\ is consistent  with \hierarchy\ when $\mu\ll\Lambda$, which is natural in the
effective field theory, and is satisfied in the geometric brane regime as well.

From the field theory perspective, the constraint \constphi\ is due to the presence of unHiggsed
magnetic quarks $q$, $\tilde q$ in vacua with $k<N_f-N_c$. In the regime of interest, the K\"ahler
potential for these quarks can be taken to be canonical, and their potential is given by \vveeffdef.

Diagonalizing the mass matrix for the bosonic and fermionic components of $q$, $\tilde q$, we
find\foot{For simplicity we restrict to the case $k=0$, in which the analysis simplifies somewhat.
Other cases are qualitatively similar.}
\eqn\mbosonq{ m_b^2=|h\mu|^2 ( |x|^2 \pm  | 1 - \epsilon x| ) ~,
~~~ ({\rm degeneracy} ~ 2 n (N_f - N_c))~,}
\eqn\mfermionq{ m_f^2=|h\mu|^2 |x|^2 ~, ~~~ ({\rm degeneracy} ~ 4 n (N_f - N_c))~,}
where $x$ is defined in  \constphi\ and
\eqn\phinx{\epsilon= \frac{h\mu_\phi}{\mu} =  \frac{ l_s \theta}{v_2} \sqrt{2\pi a}~.  }
Requiring that the scalar masses \mbosonq\ are non-tachyonic leads to the constraint
$x>1-O(\epsilon)$, in agreement with \constphi.

Thus, we see that the non-trivial K\"ahler potential for $\Phi$, \KahlerNoAngle, gives rise
to a large set of metastable vacua, some of which exist  only in the range \concorr. Note
that our results are consistent with the recent analysis of \EssigKZ. These authors showed
that taking the K\"ahler potential of $\Phi$ to be canonical, the one loop potential in the WZ
model is not sufficient for pushing the metastable vacua discussed in \GiveonEF\ to the
region \constphi, in which the unHiggsed magnetic quarks are non-tachyonic. From the point
of view of \EssigKZ, our analysis takes place at a finite value of the UV cutoff in the WZ model.
The Lagrangian includes an infinite set of non-renormalizable operators, which seem unmotivated
from the low energy point of view, but are determined in string theory. As we saw in \hierarchy,
the scales associated with these operators are in fact rather low.

\subsec{From K\"ahler to one loop}

The brane system we are studying reduces to a WZ model with
superpotential \wdef\ and canonical K\"ahler potential when the displacement $v_2$ in
figure 4 is taken to be well below $g_sl_s$. In that regime, the leading corrections to
the dynamics of $\Phi$ come from the one loop potential, which is not sufficient to produce
locally stable minima in the tachyonic branches. On the other hand, for larger $v_2$ the
dominant effect is the non-trivial K\"ahler potential and other D-terms, which in general
lead to the appearance of locally stable vacua in tachyonic branches.

It would be interesting to study the transition between the two regimes in detail by following
the dynamics as one changes the parameters of the brane model. Unfortunately, this is
difficult for reasons that were described above. In this subsection we consider a toy model,
which we believe captures the essence of the problem.

We take the potential to be a sum of classical and one loop contributions,
\eqn\vtreeloop{V=V_0+V_1~,}
where $V_0$ is the tree level potential \nearorig, and $V_1$ the Coleman-Weinberg
potential
\eqn\cw{ V_1 = \frac{1}{64\pi^2} {\rm Str} \left( m^4 \log m^2\right)~.}
The supertrace runs over the spectrum \mbosonq, \mfermionq, relevant for vacua with
$k=0$ in figure 4. We fix the parameters $h$, $\mu$, $\mu_\phi$ that enter the classical 
Lagrangian of the WZ model, and vary the UV scale $\tilde\Lambda$ that parameterizes the 
deviation from canonical K\"ahler potential.

Consider first the case $\mu_\phi\ll h\mu$. When $\tilde\Lambda$ is small, the
non-canonical K\"ahler potential \kahsmall\ is dominant, and the potential \vtreeloop\
has a local minimum at \metmin. The location of this minimum increases as $\tilde\Lambda$
increases; at the same time the size of the quadratic term in the potential for $X$,
\nearorig, decreases. At\foot{Here and below we neglect numerical factors that depend on
$N_f$, $N_c$.} $\tilde\Lambda\sim \mu/h$,
the quadratic term in the one loop potential $V_1$, \voneloop, becomes comparable to the K\"ahler
one, and as $\tilde\Lambda$ is increased further it dominates over it. In this regime, the
location of the local minimum \metmin, $\langle \Phi_n\rangle \sim \mu_\phi/h\ll\mu$, no longer
increases with $\tilde\Lambda$. Since this value is not in the regime \constphi, the corresponding
vacua are unstable to condensation of the tachyonic modes of $q$, $\tilde q$ discussed above.

As $\tilde\Lambda$ increases further, additional local minima of the potential \vtreeloop\ appear.
In particular, when $\tilde\Lambda\gg \mu^2/\mu_\phi$, one can show that the local minimum of $V_0$
\metmin\ becomes again a local minimum of the full potential \vtreeloop; the one loop correction to
$V_0$ at that point is negligible. Since the value of \metmin\ in that case is parameterically larger
than $\mu$, this vacuum is metastable.

Increasing $\tilde\Lambda$ further leads eventually to violation of the bound \negclass, after which
other contributions to the potential become important. For $\tilde\Lambda$ well above
the supersymmetric vacuum \susyvac, the contribution of the non-canonical K\"ahler potential to the
dynamics can be neglected, and the analysis reduces to that of \EssigKZ. In particular, the metastable
vacua due to the K\"ahler potential disappear in this regime.

For $\mu_\phi\sim h\mu$ or larger, the one loop contribution to \vtreeloop\ is
small for all $\tilde\Lambda$ in the range \negclass. Thus, the analysis of the classical potential
with a non-trivial K\"ahler potential is valid. For $\tilde\Lambda$ outside the range \negclass,
the metastable states disappear, as before.

Thus, we see that {\it metastable vacua in the tachyonic branches exist when the UV cutoff in the WZ
model is sufficiently small}. When the cutoff exceeds a value comparable to the scale set by the
supersymmetric vacuum, \susyvac, they disappear. This picture is in agreement with the results of
\EssigKZ\ who found no metastable states in the limit $\tilde\Lambda\to\infty$.

\newsec{Discussion}

In this paper we studied non-supersymmetric vacua of systems of intersecting $D$-branes
and $NS5$-branes in string theory. Such systems are very useful for embedding
supersymmetric gauge theory dynamics into string theory \GiveonSR, and it is natural to
ask whether they shed light on supersymmetry breaking as well.

Previous work on these systems showed that they exhibit some qualitative similarities to
the field theories discussed by ISS \IntriligatorDD, but some important differences were
noted as well. In particular, the detailed phase structure and mechanism for the
stabilization of the pseudo-moduli $\Phi$ are in general different in (renormalizable)
field theory and in string theory.

The purpose of this work was to provide a description of the string theory analysis
in terms of the light degrees of freedom. We found that
while in \IntriligatorDD\ the pseudo-moduli were stabilized by one loop effects in the
corresponding Wess-Zumino model, string theory leads naturally to a different class of
effective field theories. In addition to the superpotential \wmaggg, \wdef, one finds
a non-trivial K\"ahler potential, \KahlerNoAngle, \kahsmall, and higher order D-terms
such as \DFour. These D-terms play an important role in analyzing the vacuum structure
of the theory, and give rise to a rich landscape of metastable supersymmetry breaking vacua.

In the low energy theory the D-terms correspond to non-renormalizable operators that
require a UV completion. Such a completion is provided by string theory. It is in general
different from the UV completion proposed by ISS, in terms of an asymptotically free
Seiberg dual field theory, but since in the brane system one can continuously interpolate
between the two, many features of the phase structure are qualitatively similar.

We calculated the K\"ahler potential of the pseudo-moduli in the approximation where the
$D$-branes giving rise to them can be treated as local probes of the geometry of the extra
dimensions, and found that in that approximation the K\"ahler potential is determined by
the geometry, \KahlerNoAngle. The fact that the metric on a supersymmetric
moduli space of $D$-branes moving in a non-trivial geometry depends simply on that geometry
is well known. Here we found that this property persists to non-supersymmetric (pseudo)
moduli spaces. This was established to leading order in the supersymmetry breaking
parameter, but is likely to be true more generally.

There are a number of natural extensions of this work that are worth pursuing. In the
supersymmetric case, the structures seen in type IIA brane systems of the sort considered here
have IIB counterparts in terms of $D$-branes wrapping small cycles on Calabi-Yau manifolds. It
would be interesting to extend these results to non-supersymmetric systems, and in particular
calculate the K\"ahler potential and higher D-terms in that case.

It would also be interesting to consider the time-dependent dynamics of systems of the sort considered
here, that might be relevant for early universe cosmology. If the $NS$-brane that produces the
non-canonical D-terms is in motion with respect to the $D$-branes on which the pseudo-moduli live,
the potential changes with time. The system could be trapped for a while in a metastable minimum, and 
then decay when the fivebrane reaches a critical distance from the $D$-branes.
This might give interesting models of inflation in string theory.

One can try to use models of the sort described here as hidden sector models for
supersymmetry breaking in nature. In particular, the metastable vacua that we found in the tachyonic
branches, which owe their existence to the non-trivial K\"ahler potential, appear to be particularly
promising candidates for phenomenology, due to the large breaking of R-symmetry in them.

As an example, one can consider the brane configuration of figure 4 in the metastable
state with $k=0$, and arbitrary $n$. In this vacuum, the system has a global symmetry $U(n)\times
U(N_f-n)$. It is natural to embed the gauge group of the MSSM in the $U(n)$ factor. The
pseudo-modulus $\Phi_n$ is the SUSY breaking chiral spurion, and the magnetic quarks $q$,
$\tilde q$ are the messengers. Standard gauge mediation (see \eg\ \GiudiceBP\ for a review) leads
to comparable gaugino and sfermion masses of the order of $\alpha\mu^2/X_{\rm min}$. One can
choose the parameters of the model such that these masses are around the weak scale.

\bigskip
\noindent{\bf Acknowledgements:} We thank O. Lunin and G. Torroba
for discussions. This work is supported in part by the BSF --
American-Israel Bi-National Science Foundation. AG is supported
in part by a center of excellence supported by the Israel Science
Foundation (grant number 1468/06), DIP grant H.52, and the Einstein
Center at the Hebrew University. DK is supported in part by DOE
grant DE-FG02-90ER40560 and the National Science Foundation under
Grant 0529954. JM is supported in part by a Ledley fellowship. AR
is supported in part by a Department of Education GAANN fellowship
P200A060226. AG thanks the EFI at the University of Chicago for
hospitality during the course of this work.

\appendix{A}{Fermionic terms in the low energy effective Lagrangian}

In section 2.2 we calculated the leading terms in the low energy effective Lagrangian of the bosonic (bottom)
component of the chiral superfield $\Phi$. In this appendix we show that the fermionic terms are consistent
with \lkinfin, with the K\"ahler potential \KahlerNoAngle.

Expanding \lkinfin\ in components (and restricting again to a single $D4$-brane, for simplicity) gives \WessCP:
\eqn\BosFermi{\CL_2 = -\KK_{\Phi\bar\Phi} \left(|\del_\mu\Phi|^2 + i\bar\psi \slash D_\mu \psi \right) + \cdots}
where $D_\mu\psi  =\partial_\mu\psi + \del_\mu \Phi\,\Gamma^\Phi_{\Phi\Phi}\psi$, with
$\Gamma^\Phi_{\Phi\Phi} = \del_\Phi H(|\Phi|)$. The  supersymmetry transformation is
\eqn\PhiSUSY{\delta_\zeta \Phi = \sqrt{2} \zeta \psi~.}
To compute the fermionic term in \BosFermi\ for the brane system, we start with the DBI action for the $D4$-brane,
keeping terms quadratic in the fermions \refs{\AganagicPE\BergshoeffTU-\MarolfVF}:
\eqn\DBIFermion{\LL_{\rm D4~fermi} = \frac{i}{2} T_4 \int d^5\sigma e^{-\varphi} \sqrt{-{\rm det}P(G+B)} ~\bar \Theta (1+\wt \Gamma_{D_4}) (P(\Gamma^a D_a) - \Delta) \Theta~.}
Here $\Theta$ is the thirty-two component Majorana spinor, $D_A = \nabla_A +\frac{1}{4} H_{ABC}\Gamma^{BC}\bar\Gamma$
is the torsional connection, and $\Delta = \half \left(\Gamma^A \del_A \varphi + \frac{1}{6} H_{ABC}\Gamma^{ABC}\bar\Gamma\right)$.
The matrix $\wt\Gamma_{D4} = \frac{1}{5!} \e^{a_1\ldots a_5} \Gamma_{a_1\ldots a_5} \bar\Gamma$ is the kappa
symmetry projector, and $\bar{\Gamma}$ is the usual chirality matrix in ten dimensions. $\sigma^a$, $a=0,\ldots,4$ are
worldvolume coordinates, while $A,B$ are ten-dimensional tangent space indices.

The first step is to extract the superpartner $\psi$ of $\Phi$ from the thirty two component spinor $\Theta$.
The supersymmmetry transformations of the transverse scalars $X^I(=v,w,x^7)$ take the form
\eqn\DBISUSY{\delta X^I = -\frac{i}{2}\bar\epsilon \Gamma^I \Theta~,}
The configuration we study (figure 1 with $v_2=0$) preserves four supercharges. The corresponding supersymmetry
 parameter, $\epsilon$, is determined by the projection conditions
\eqn\BPS{\eqalign{&\Gamma_{NS'} \epsilon = \Gamma^0\ldots\Gamma^3 \Gamma^{w_1}\Gamma^{w_2}\epsilon = \epsilon~,\cr
& \Gamma_{D6}\epsilon =  \Gamma^0\ldots\Gamma^3\Gamma^{w_1}\Gamma^{w_2}\Gamma^7\epsilon = \epsilon~,\cr
& \Gamma_{D4} \epsilon =\Gamma^0\ldots\Gamma^3\Gamma^y\bar\Gamma\epsilon=\epsilon~.}}
We pick a basis for the gamma matrices compatible with the symmetries of the brane configuration
\eqn\GammaBasis{\eqalign{&\Gamma^\mu = \sigma^2 \otimes \gamma^\mu\otimes 1_2 \otimes 1_2~,\cr
&\Gamma^{v_1,v_2,y} = \sigma^1\otimes 1_4 \otimes \sigma^{1,2,3} \otimes 1_2~,\cr
&\Gamma^{w_1,w_2,7} = \sigma^3\otimes 1_4 \otimes  1_2 \otimes \sigma^{1,2,3}.~\cr}}
The projection conditions imply $\epsilon$ is of the form
\eqn\SpinorOne{ \epsilon = \chi_+ \otimes \epsilon_R \otimes \chi_+ \otimes \chi_+ + {\rm c.c.}~,}
where the complex conjugate is determined by the Majorana condition $\epsilon^* = B \epsilon$
($B$ is the product of all the real gamma matrices). $\chi_\pm$ is a constant two-dimensional spinor
satisfying $\sigma^3\chi_\pm = \pm\chi_\pm$, and $\epsilon_R$ is a right-handed Weyl spinor in
the $\IR^{3,1}$ worldvolume of the brane configuration.

Recalling the preserved supersymmetries of the DBI action are given by \DBISUSY, using \SpinorOne\
and setting $I=w$ in \DBISUSY, we see that the superpartner of $\Phi$ has the following embedding in $\Theta$:
\eqn\ThetaDBI{ \Theta = \chi_- \otimes \psi_L \otimes \chi_+ \otimes \chi_-+{\rm c.c.}}
To relate this to the conventional form for a Weyl spinor in four-dimensions, we choose a Weyl basis such that
$\epsilon_R ={1\over\sqrt2} (0,\bar\zeta_{\dot\alpha})^T$ and $\psi_L ={1\over\sqrt2} (\wt \psi^\alpha, 0)^T$.
Then, using \cannorm, \DBISUSY\ gives us the supersymmetry variation for $\Phi$
\eqn\DBISUSYTwo{\delta \Phi = \sqrt{\frac{2T_4 \Delta y}{g_s }} H(|\Phi|)^{-\half}\zeta \wt \psi~,}
where the $H^{-1/2}$ comes from the vielbein in $\Gamma^w$. Comparing to \PhiSUSY\ we identify the canonically
normalized superpartner of $\Phi$,
\eqn\fermionRenorm{\psi = \sqrt{\frac{T_4 \Delta y}{g_s }} H(|\Phi|)^{-\half} \wt\psi~.}
Finally, plugging \fermionRenorm, \ThetaDBI\ into \DBIFermion, and evaluating on the CHS background \chs, we find the expected result \BosFermi.

\listrefs
\end

It will be useful in the following to introduce two more dimensionless quantities
\eqn\deltay{  y = \frac{y_{NS}}{l_s} ~, ~~~ \delta = \frac{2\pi h \mu}{m_s} = \sqrt{ \frac{ 2\pi v_2}{\Delta y}} = \sqrt{2\pi a} ~,}
in terms of which  becomes
\eqn\comcontwo{ \frac{\delta}{y} < \frac{ \epsilon y^3}{\delta} \ll 1~.}
We also note that the K\"ahler potential takes the form $H(x) = 1 + (y^2 + \delta^2 x^2)^{-1}$ in these parameters.

Following the discussion at the end of section 2, the DBI regime corresponds to finite $a$, and thus finite $\delta$.  In this regime K\"ahler corrections dominate over one-loop effects.  Therefore the above analysis is appropriate for finite $\delta$, and by choosing $\epsilon$ and $y$ according to \comcontwo, it demonstrates the existence of a metastable miminum in the DBI regime of the tachyonic branch.  Note that as $y_{NS} \rightarrow \infty$, the metastable minimum disappears as expected--as we remove the $NS$-brane, the K\"ahler potential becomes canonical.  On the other hand, as $\delta \rightarrow 0$, field theory effects become important, and the contribution of the one-loop potential eventually dominates the contribution of the K\"ahler potential.  In this limit the analysis of \EssigKZ\ shows that there is no metastable minimum.  In the following we will observe the crossover by taking into account both contributions and varying the parameters.

For simplicity, we take $\Phi_n$ to be proportional to the identity,
and introduce two dimensionless quantities $x,\epsilon$:
\
Next we consider the effective potential for $x$, setting the background value of the quarks to zero.

The tree level potential is
\eqn\Vtree{ V_{tree}(x) = {n\over H(x)}|h^2\mu^4||1-\epsilon x|^2 ~.}
The quark fluctuations that couple to the pseudomodulus $x$ may develop a negative mass-squared.  This is the manifestation in field theory of the open string tachyon discussed above.  We find that the bosonic and fermionic mass-squared eigenvalues of these superfields are

Since the nontrivial K\"ahler potential corresponds to non-renormalizable operators from the low energy field theory point of view, one expects the crossover region to be at the cut-off scale.  Therefore we will compute the one-loop potential with a finite cut-off, $\Lambda_{\rm cutoff}$, which modifies the standard Coleman-Weinberg result:
\eqn\cw{ V_{1-loop} = \frac{1}{64\pi^2} \rm{Str} \left( m^4 \log{\left[ \frac{m^2}{\Lambda_{\rm cutoff}^2} \right]} + (\Lambda_{\rm cutoff}^4 - m^4) \log{\left[ 1 + \frac{m^2}{\Lambda_{\rm cutoff}^2} \right]} \right) ~.}
Here we have dropped terms from the integral $\frac12\int_{0}^{\Lambda_{\rm cutoff}} {d^4p\over (2\pi)^4}\log{(p^2 + m^2)}$ that either cancel in the supertrace or renormalize the tree
level potential.  Only quark fluctuations that couple to $\Phi_n$ contribute to the supertrace.  They have mass eigenvalues given by \mbosonq, \mfermionq.  Due to the nontrivial K\"ahler potential, fluctuations of $\Phi_n$ also contribute to the supertrace.   However, since these contributions are proportional to derivatives of the K\"ahler potential, they are down by factors of $l_s/y_{NS} = 1/y$ relative to the quark contributions.  They can be easily computed but are completely negligible.  It is natural to take the field theory cut-off to be the string scale.  Specifically we choose $\Lambda_{\rm cutoff} = m_s/2\pi$.  Denoting the three mass eigenvalues by by $m^2 = |h\mu|^2 \lambda^2$, we find that the effective potential to one loop is
\eqn\Veff{ \frac{1}{n |h^2\mu^4|} V_{eff}(x)  = H(x)^{-1} | 1 - \epsilon x|^2
+ \frac{(N_f - N_c) |h|^2}{32 \pi^2} \VV(x) ~,}
where
\eqn\Veffb{ \eqalign{\VV &= \lambda_{+}^4 \log{\left[ \frac{\delta^2 \lambda_{+}^2}{ 1+ \delta^2 \lambda_{+}^2} \right]} + \lambda_{-}^4 \log{\left[ \frac{\delta^2 \lambda_{-}^2}{ 1+ \delta^2 \lambda_{-}^2} \right]} - 2 \lambda_{f}^4 \log{\left[ \frac{\delta^2 \lambda_{f}^2}{ 1+ \delta^2 \lambda_{f}^2} \right]}  +\cr
&~~~~+ \frac{1}{\delta^4} \log{\left[ 1 - \frac{\delta^2 H^{-2} | 1 - \epsilon x|^2 }{( 1+ \delta^2 |x|^2)^2} \right]} ~.}}

We can estimate when the one-loop term ceases to be a negligible contribution to the tree-level term.  We must work to order $1/y^4$ in order to have a nontrivial K\"ahler potential, while from \comcontwo\ it follows that $\epsilon$ must be at least $1/y^4$.  Therefore in the tree-level term we can consistently work to order $1/y^4$ and order $\epsilon$, precisely what was done in \nearorig.  In the one-loop term we initially make no assumptions about $\delta$ or $x$, but work to zeroth order in $\epsilon, 1/y$.  After differentiating there are some simplifications and one can show
\eqn\DVeffb{\partial_x \VV = 4 x^3 \log{ \left[  \frac{(x^4-1)(1+2 \delta^2 x^2)^2}{x^4 ((1 + \delta^2 x^2)^2 - \delta^4)} \right] } + 4 x \log{ \left[ \frac{ (x^2 +1)(1 + \delta^2 (x^2 -1)) }{ (x^2 -1)(1 + \delta^2(x^2 + 1))} \right] } + \OO(y^{-2})~.}
To proceed further we expand in $1/x$.  This may seem ill-advised since we expect the metastable vacuum to be near the tachyon region $x \sim \OO(1)$.  On the other hand one finds, remarkably, that the leading order term is $1/x^5$.  Adding this sum to $\partial_x V_{tree}$ and setting the result to zero yields
\eqn\oneoverx{ \frac{2\delta^2}{y^4} x - 2 \epsilon + \frac{(N_f-N_c) h^2 }{8 \pi^2 \delta^4 x^5} \left( 1 - \frac{2}{\delta^2 x^2} + \OO(y^{-2}, x^{-4}) \right )  = 0~.}
The tree-level solution, $x = x_0$, given in \metmin, and leading order one-loop corrections to that solution, $\delta x$, are
\eqn\xsolapp{ x_0 = \frac{\epsilon y^4}{\delta^2}~, ~~~ \delta x = \frac{(N_f-N_c) h^2 }{8 \pi^2 \delta^4 x_{0}^5}   ~.}
For finite $\delta$, the correction can be small for values of parameters satisfying \comcontwo.  However, as $\delta$ is decreased the correction will eventually become large, invalidating the analysis.

In the sequence of plots in figure 5, we see how decreasing $1/y^2$ while holding all other parameters fixed destroys the metastable vacuum, as expected.

\ifig\loc{Varying $y=y_{NS}/l_s$ over a range of values with $h=0.3$, $\epsilon/h = \mu_\phi/\mu = 9\times10^{-5}$, $\delta = 2\pi h\mu / m_s = 0.6$ fixed. We see the metastable vacuum disappears for sufficiently large $y_{NS}$, while as $y_{NS}$ approaches the v-axis the metastable vacuum is pushed towards the tachyonic region.}
{\epsfxsize6.0In\epsfbox{sec3_plot1.eps}}

In the next sequence of plots, figure 6, we decrease $\delta$ while holding all other parameters fixed.  This corresponds to descending into the low energy effective field theory regime.  Again, as expected, the metastable vacuum gets washed out.  This demonstrates that the metastable vacuum is due to cut-off scale physics.

One should not be alarmed by this statement.  An analogous effect was observed in the Nambu-Jona-Lasinio model of chiral symmetry breaking in QCD \NambuTP.  There it was shown that by adding a non-renormalizable four-fermi coupling to the Lagrangian and working with a finite cut-off, one could obtain a nonzero quark condensate.  The strength of the four-fermi coupling is order one in units of the cut-off; however, the value of the condensate --
the physical quantity that can be measured -- is well below the cut-off, thus there is reason to believe the qualitative features of the analysis.  Here, as well,  there is a scale which is well below the cut-off.   This is the scale of fluctuations around the metastable minimum, and from \nearorig, we see that this mass scale is \massphi.  The feature that really distinguishes this model, however, is that the cut-off scale physics can be determined by a calculable UV completion.

\ifig\loc{Varying $\mu/m_s$ over a range of values with $h=0.3$, $\epsilon/h=9\times10^{-3}$ and $y=17$ fixed. The metastable minimum only appears when $\mu \sim m_s$ -- that is, at the cut-off scale, and vanishes for $\mu \ll m_s$, which is well within the field theory regime. }
{\epsfxsize6.0in\epsfbox{sec3_plot2.eps}}

When $k>0$ there is a new set of fluctuations that contribute to the one-loop potential, which can be found in \EssigKZ. It is clear from the brane perspective, however, that these should not qualitatively modify the results. One should be careful in computing the Coleman-Weinberg 1-loop potential in the field theory regime. Some of the fluctuations correspond to Nambu-Goldstone bosons, and it naively looks as though their masses are tachyonic. This effect is not physical, and is a manifestation of the origin of field space (our semi-classical starting point) not being the vacuum. Thus, the tachyonic mass is expected to be removed at 1-loop, and consequently should be ignored when computing the $STr$ in the Coleman-Weinberg potential.